\documentclass[9pt,twocolumn,twoside]{osajnl}

\usepackage[T1]{fontenc}
\usepackage{babel}
\usepackage{verbatim}
\usepackage{amsmath}
\usepackage{breakurl}

\usepackage{graphicx} %to include PDFs
\usepackage{pdfpages}

\makeatletter

\providecommand{\tabularnewline}{\\}

\mathcode"65"65 %Make e upright
\usepackage{upgreek} %Make mu upright for um
\usepackage{textcomp}  %copyright symbol

\usepackage{amsmath} %Allows for 'subequations'
\let\int\relax

\usepackage{tikz}                       % Für pgfplots benötigt
\usepackage{pgfplots}                   % Für schöne Plots
\usepgfplotslibrary{external}
\usetikzlibrary{decorations}

\usepackage{soul}
\setulcolor{red}
\setul{-0em}{1.5pt}
\newcommand{\redline}[1]{{#1}} %remove redline

\makeatletter
\renewcommand{\fnum@table}{TABLE~\thetable}
\renewcommand{\fnum@figure}{FIG.~\thefigure}
\makeatother

\renewcommand{\thetable}{\Roman{table}}

\journal{ol} % Choose journal (ao, aop, josaa, josab, ol)

\setboolean{shortarticle}{true} 
\ifthenelse{\boolean{shortarticle}}{\colorlet{color2}{color2b}}{\colorlet{color2}{color2}} % Automatically switches colors for short articles

\title{Beyond the perturbative description of the nonlinear optical response of low-index materials}

\author[1,*]{Orad Reshef}
\author[1]{Enno Giese}
\author[1]{M. Zahirul Alam}
\author[2]{Israel De Leon}
\author[1]{Jeremy Upham}
\author[1,3]{Robert W. Boyd}

\affil[1]{Department of Physics, University of Ottawa, 25 Templeton Street,
Ottawa, Ontario K1N 6N5, Canada}
\affil[2]{School of Engineering and Sciences, Tecnológico de Monterrey, Monterrey,
Nuevo León 64849, Mexico}
\affil[3]{Institute of Optics and Department of Physics and Astronomy, University
of Rochester, Rochester, NY 14627, USA }
\affil[*]{e-mail: orad@reshef.ca}

\dates{Compiled \today}

\ociscodes{(190.0190) Nonlinear optics; (190.3270) Kerr effect.}

\doi{\url{http://dx.doi.org/10.1364/ao.XX.XXXXXX}}

\begin{abstract}
We show that standard approximations in nonlinear optics are violated for situations involving a small value of the linear refractive index. Consequently, the conventional equation for the intensity-dependent refractive index, $n(I) = n_0 + n_2I$, becomes inapplicable in epsilon-near-zero and low-index media, even in the presence of only third-order effects. For the particular case of indium tin oxide, we find that the $\chi^{(3)}$, $\chi^{(5)}$ and $\chi^{(7)}$ contributions to refraction eclipse the linear term; thus, the nonlinear response can no longer be interpreted as a perturbation in these materials. Although the response is non-perturbative, we find no evidence that the power series expansion of the material polarization diverges.
\end{abstract}

\setboolean{displaycopyright}{true}

\begin{document}
% Add these next 3 lines and the pdfpages to include the copyright page
\null %Necessary if \includepdf is right after \begin{document}
\includepdf{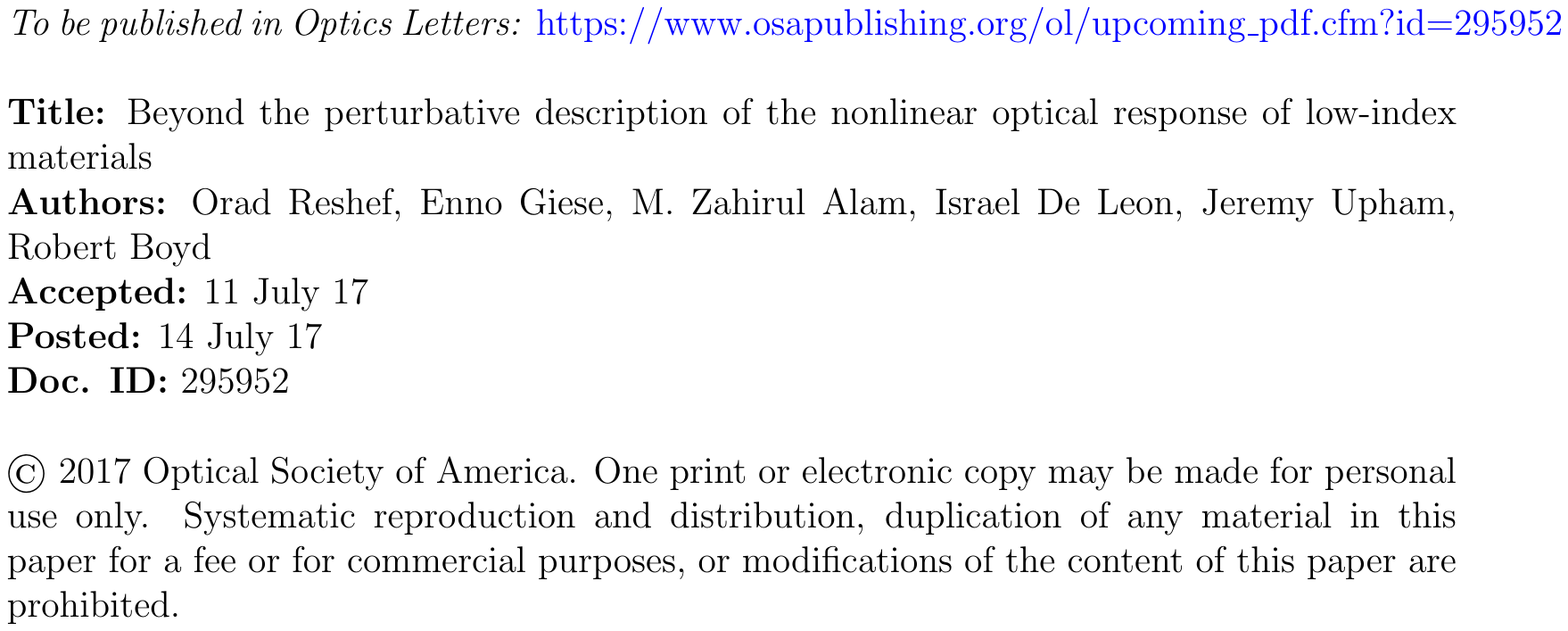}
\setcounter{page}{1}

\maketitle
\thispagestyle{fancy}

\ifthenelse{\boolean{shortarticle}}{\ifthenelse{\boolean{singlecolumn}}{\abscontentformatted}{\abscontent}}{}

%\section{Introduction}
Nonlinear optical effects are crucial to many applications in photonics, performing
essential functions in lasing, frequency conversion, and entangled-photon
generation, among others~\cite{Siegman1986,Kwiat1995,Kippenberg2011,Boyd2008}.
Due to the small intrinsic nonlinearities of common photonic materials,
there has long been the desire to increase optical nonlinearities
in order to increase conversion efficiencies, miniaturize device footprints
and reduce power requirements in optical devices. Much of the recent
work has been towards enhancing nonlinearities by structuring materials,
for example using high-$Q$ micro-cavities~\cite{Heebner1999,Bravo-Abad2007,Soljacic2004}
or photonic crystals with slow-light propagation~\cite{Soljacic2004,Hamachi2009}.

Recently, a class of low-index materials called epsilon-near-zero
(ENZ) materials, whose real part of the electric permittivity $\epsilon'$
vanishes at a certain wavelength, has emerged as a promising platform
to achieve unprecedentedly large nonlinear responses~\cite{Alam2016,Caspani2016,Capretti2015}.
For example, in indium tin oxide (ITO), the nonlinear contribution
to the index of refraction $\Delta n$ has achieved a value of $0.72$~\cite{Alam2016}.
This value is considerably larger than what has been achieved in highly
nonlinear chalcogenide glasses ($\Delta n\approx10^{-6}$)~\cite{Harbold2002, Slusher2004,Eggleton2011},
and could enable all-optical switching in a propagation length smaller than
a single wavelength. With recent developments in the integration of
zero-index metamaterials, whose zero-refractive-index wavelength can
be arbitrarily selected to suit the application \cite{Huang2011,Moitra2013,Li2015a,Kita2017,Vulis2016arXiv},
it has become critical to conduct an in-depth investigation
of the nonlinear optical response of low-index media.

The recently demonstrated magnitude of nonlinear responses of ENZ
materials is paradigm-shifting, and questions certain established
fundamental assumptions in the field of nonlinear optics. For example, in a recent publication on the nonlinearity of aluminum-doped zinc oxide (AZO), the authors claim that ``the ENZ nonlinearity in AZO [is] in a regime where the approximation of expanding the material polarization in a power series breaks down''~\cite{Caspani2016}.
%A point has been raised on whether the power series expansion description
%of nonlinear optical effects requires reformulation in these materials~\cite{Alam2016,Caspani2016}.
Here, we theoretically and experimentally explore the consequences
of a vanishingly small permittivity on the nonlinear
optical response. %In \textbf{Section~\ref{sec:intensity-dependent-index}}, we rederive the intensity-dependent index of refraction for highly-nonlinear low-index materials and compare our theory to experimental results. In \textbf{Section~\ref{sec:higher-order}}, we show that for the nonlinear responses in ENZ materials, the nonlinear polarization can no longer be considered as a perturbation term; however, despite speculations to the contrary, the power series description of the nonlinear polarization appears to remain valid. 

%\section{Intensity-dependent refractive index due to third-order effects}
%\label{sec:intensity-dependent-index}
We begin by deriving an expression for the intensity-dependent index
of refraction caused solely by the third-order nonlinear susceptibility
$\chi^{(3)}$. For simplicity, we assume a centrosymmetric material, neglect the tensor nature of the susceptibility, as well as
material magnetic responses. This set of assumptions is reasonable for
most nonlinear optical materials~\cite{Boyd2008}. 

To lowest nonlinear order, the polarization of a material 
illuminated by a monochromatic laser field is described as:
\begin{align}
P^{\textrm{TOT}} & =P+P^{\textrm{NL}}=\epsilon_{0}E\left[\chi^{(1)}+3\chi^{(3)}|E|^{2}\right].
\end{align}
Here, $E$ is the complex amplitude of the applied electric field
and $\chi^{(1)}\equiv\epsilon^{(1)}-1$ corresponds to the linear
response of the material, with $\epsilon^{(1)}$ being the linear
relative permittivity. The relative permittivity including only
the $\chi^{(3)}$ nonlinearity is thus
\begin{equation}
\epsilon=\epsilon^{(1)}+3\chi^{(3)}|E|^{2}.\label{eq:epsilon}
\end{equation}

Since all of these quantities may be complex, we define the complex
relative permittivity as $\epsilon=\epsilon'+i\epsilon''$ and the
complex refractive index as $n=n'+in''$, where a single prime denotes
the real part, and the double prime the imaginary part, respectively.
These two quantities are related by~\cite{DelCoso2004}
\begin{equation}
n=\sqrt{\epsilon}=\sqrt{\epsilon^{(1)}+3\chi^{(3)}|E|^{2}}.\label{eq:intensity-dependent index withchi}
\end{equation}

%begin{subequations}
%\begin{align}
%n' & =\sqrt{\frac{|\epsilon|+\epsilon'}{2}},\,\,\,\,\,\,\,\,\,\,n''=\sqrt{\frac{|\epsilon|-\epsilon'}{2}}\label{eq:n_epsilon_a}\\
%\epsilon' & =n'^{2}-n''^{2},\,\,\,\,\,\,\,\,\,\,\epsilon''=2n'n''\label{eq:n_epsilon_b}
%\end{align}
%\end{subequations}
Together, these equations can be used to obtain the complex, intensity-dependent
index of refraction $n$ due to third-order contributions. We find
that
\begin{align}
n & =\sqrt{n_{0}^{2}+2n_{0}n_{2}I},\label{eq:intensity-dependent index}
\end{align}
where we take $n_{0}=\sqrt{\epsilon^{(1)}}$ to be the linear refractive
index, $I$ to be the optical field intensity
\begin{equation}
I=2\textrm{Re}(n_{0})\epsilon_{0}c|E|^{2},\label{eq:I_definition}
\end{equation}
and we introduce the standard definition for the nonlinear index of
refraction~\cite{DelCoso2004,Boyd2008}
\begin{equation}
n_{2}=\frac{3\chi^{(3)}}{4n_{0}\textrm{Re}(n_{0})\epsilon_{0}c}.\label{eq:n2_definition}
\end{equation}

In order to obtain a simpler relation for $n$, Eq.~(\ref{eq:intensity-dependent index})
is usually expanded in a power series under the assumption that $|2n_{2}I/n_{0}|\ll1$~\cite{Boyd2008},
yielding
\begin{align}
n & =n_{0}\sqrt{1+2\frac{n_{2}I}{n_{0}}}\approx n_{0}\Bigg[1+\frac{1}{2}\left(2\frac{n_{2}I}{n_{0}}\right)+\ldots\Bigg].\label{eq:n_power_series}
\end{align}
In most materials, $|2n_{2}I/n_{0}|$ is very small so that only the
lowest order correction term is kept, resulting in the intensity-dependent refractive
index being widely defined as
\begin{equation}
n=n_{0}+n_{2}I.\label{eq:n_smallI}
\end{equation}
Hence, the change of the refractive index due to the nonlinearity
is $\Delta n=n-n_{0}\approx n_{2}I$.

At this point, we pause to address a few concerns with this derivation
when considering a vanishingly small index. First, in an
ENZ material, $\Delta n/n_{0}$ can be larger than unity (\emph{e.g.,
}in Al-doped ZnO, this ratio has been shown to equal 4.4~\cite{Caspani2016}).
In this case, the assumption that permits the power series expansion
of Eq.~(\ref{eq:n_power_series}) and leads to Eq.~(\ref{eq:n_smallI})
is violated. Therefore, this power series strictly\emph{ does not
converge} and Eq.~(\ref{eq:n_smallI}) is not a valid approximation
of the intensity-dependent index of refraction.

Secondly, Eq.~(\ref{eq:intensity-dependent index}) reveals an issue
that is not immediately apparent from Eq.~(\ref{eq:n_smallI}). As
$|n_{0}|\rightarrow0$, $n$ approaches zero as well, appearing to eliminate all refraction,
including any nonlinearities. This conflict in fact also exists within
Eq.~(\ref{eq:n_smallI}) \textemdash{} as $|n_{0}|\rightarrow0$,
the optical field intensity vanishes, while simultaneously, $n_{2}\rightarrow\infty$,
leaving their product \emph{$(n_{2}I)$} in Eq.~(\ref{eq:n_smallI}) seemingly undefined. Note
that $n_{0}$ is only introduced in the nonlinear contribution to
Eq.~(\ref{eq:intensity-dependent index}) in order to obtain Eq.~(\ref{eq:n_smallI})
in the appropriate limit, and it is this factor of $n_{0}$ that leads
to the ostensible divergence of $n_{2}$ for low-index materials. No artificial effects of this kind appear
when we phrase the nonlinear optical response purely in terms of the
susceptibility and the electric field.
$\chi^{(3)}|E|^{2}$ remains a robust measure of the nonlinear response
even when $n_{2}$ and $I$ take on exceptional values.

Third, it is confounding to accurately interpret what it means to
have an intensity that is identically zero when $\textrm{Re}(n_{0})$
vanishes in Eq.~(\ref{eq:I_definition}). When employing the optical
field intensity instead of the complex field amplitude, we also need
to address whether $n_{0}$ or $n$ should be used in its definition,
which is typically not an important consideration when $\Delta n\ll1$. 

We must conclude that it does not seem beneficial to introduce $n_{2}$
or the nonlinear index of refraction as defined in Eq.~(\ref{eq:n_smallI})
in the context of low-index materials. In order to avoid these issues,
we posit that it is preferable to use the intensity-dependent index
of refraction as defined in Eq.~(\ref{eq:intensity-dependent index withchi}),
with the square root, the nonlinear susceptibility and the electric
field amplitude directly. %This form also readily extends to higher-order nonlinearities
%with the addition of terms of order $|E|^{2j}$. 
Though this equation is present in standard textbooks on nonlinear optics~\mbox{\cite{Boyd2008}}, it appears only as a step in the derivation for Eq.~(\mbox{\ref{eq:n_smallI}}). Here, we have identified the first case where its use becomes necessary to model the optical response.

We demonstrate the significance of these insights by considering experimental
results. Except where explicitly stated, we focus our discussion on
the single wavelength where the linear permittivity $\textrm{Re}(\epsilon^{(1)})=0$,
which typically corresponds to the wavelength where the index is at
its lowest, and $\Delta n$ has been shown to attain its peak value~\cite{Alam2016}.
In the experiment, originally reported in Ref.~\cite{Alam2016},
the transmission and reflection are measured through a thin film as
a function of intensity. Since the index change is so large, the Fresnel
coefficients at the boundaries of the film change dramatically, which
allows for the detection of measurable changes in these quantities
as a function of intensity. We then extract both the real and imaginary
parts of the refractive index from the measured transmission and reflection
using a transfer-matrix method~\cite{Yariv2007}. The measurements
are performed on a 310-nm-thick film of ITO using 150~fs pulses at
an oblique angle of 30$^{\circ}$. The linear permittivity of ITO
has a zero-crossing at $\lambda=1240\,\textrm{nm}$, and a correspondingly
large nonlinearity at this wavelength~\cite{Alam2016}. However,
it does not exhibit a true instantaneous Kerr nonlinearity as the
material exhibits a finite optical recovery time~\cite{Rotenberg2007,Conforti2012}. \redline{Additionally, extracting nonlinear coefficients using this method will be highly dependent on external factors other than the microscopic properties of the material. For these reasons, we denote the nonlinear orders of the susceptibility as \emph{effective} susceptibilities and } restrict our discussion to a fixed pulse width, \redline{wavelength and angle of incidence.}
We begin by examining input intensities up to 50~GW/cm$^{2}$, above
which higher-order nonlinearities begin to make significant contributions
to refraction.

\begin{figure}[htb]
\centering{}
\includegraphics[width=0.9\columnwidth]{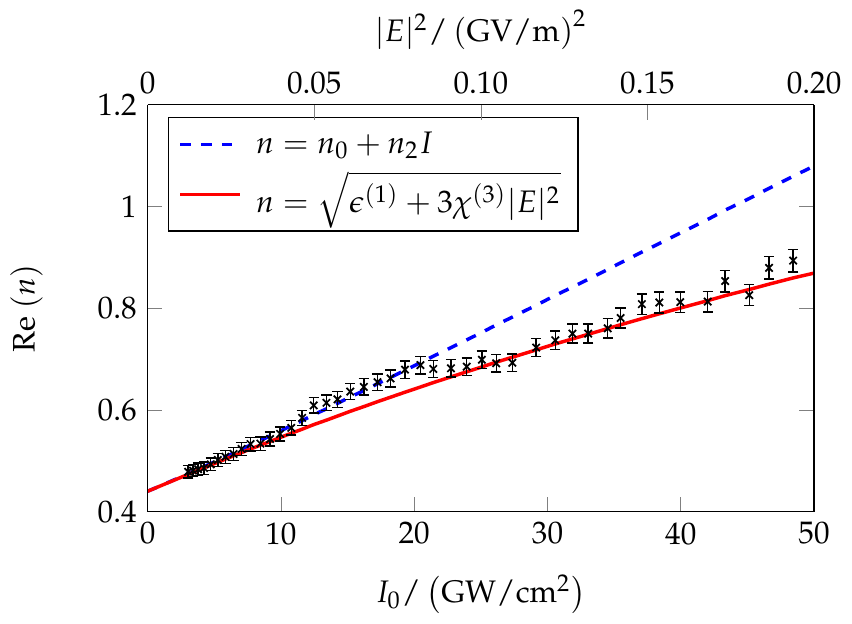}
\caption{The intensity-dependent index of refraction of ITO at $\lambda=1240\,\textrm{nm}$,
where the real part of the linear permittivity $\epsilon^{(1)}$ vanishes.
Equation~(\ref{eq:n_smallI}) performs poorly at describing the refractive
index at most intensities (dashed blue line). Using Eq.~(\ref{eq:intensity-dependent index withchi}),
we obtain much-improved agreement with the measurement without additional
fit parameters (red line). \redline{$I_0$ and $E$ refer to the free-space intensity and the corresponding field magnitude for a plane wave inside the material, respectively.}}
\label{Fig: n vs I from Science paper}
\end{figure}

We present the outcome of this measurement in Fig.~\ref{Fig: n vs I from Science paper}
as a function of the free-space incident pulse intensity \redline{$I_0$}. We also
plot Eq.~(\ref{eq:n_smallI}) for comparison, using values extracted
from independent ellipsometry and Z-scan measurements ($\textrm{Re\,}(n_{0})=0.44$
and $\textrm{Re\,}(n_{2})=\nobreak0.016\textrm{\,cm}^{2}/\textrm{GW}$)~\cite{Sheik-Bahae1990,Alam2016}.
This equation provides an adequate estimate of the index at low intensities
but quickly fails to describe the refractive index as the intensity
increases. In comparison, we plot Eq.~(\ref{eq:intensity-dependent index withchi})
using these same material values. The resulting curve follows the
measured refractive index much more accurately. We stress that the
form of this curve is due solely to the square-root nature of Eq.~(\ref{eq:intensity-dependent index withchi});
it is not caused by any absorption-based saturation effects or higher-order
contributions to the nonlinear susceptibility. Recall that this form
of the equation has been derived assuming only third-order contributions
to the nonlinear polarization. \redline{Our treatment is different from a polynomial fit to the refractive index, since it preserves the original definition of $n_2$ and describes some of the nonlinear behavior even in lowest order.} At higher input intensities, the curve
begins to deviate significantly from the measured values due to the
emergence of higher-order nonlinear effects. We discuss the contribution
of these nonlinearities to refraction in the following section.

%\section{Contributions to refraction from higher-order nonlinearities}\label{sec:higher-order}
The nonlinear polarization $P^{\textrm{NL}}(E)$ can be defined to
be a complex function of the electric field amplitude. Its explicit
form may depend highly on the experimental realization and the microscopic
model that describes the material. Therefore, its analytical form
might not be accessible at all. In the present context, we are content with expanding the nonlinear
polarization in a power series and describing the interaction by its
macroscopic properties. Thus, our method can be applied even if there
exists no good microscopic model for the material response. %As a consequence,
%this treatment is equally applicable to other highly nonlinear pure
%Kerr-type materials with instantaneous responses. 

For a single-beam input, we represent the material polarization with
the following power series
\begin{align}
P^{\textrm{TOT}}(E) & =\epsilon_{0}E\sum_{j\,\textrm{odd}}^{\infty}c_{j}\chi^{(j)}|E|^{j-1},\label{eq:PNL power series}
\end{align}
where $c_{n}$ is a degeneracy factor~~\cite{Boyd2008}. We have included only odd orders of $\chi^{(j)}$ because only
those contribute to refraction. 

%\begin{equation}
%c_{j}\equiv\binom{j}{(j-1)/2}
%\end{equation}

\begin{figure}[htb]
\includegraphics[width=0.9\columnwidth]{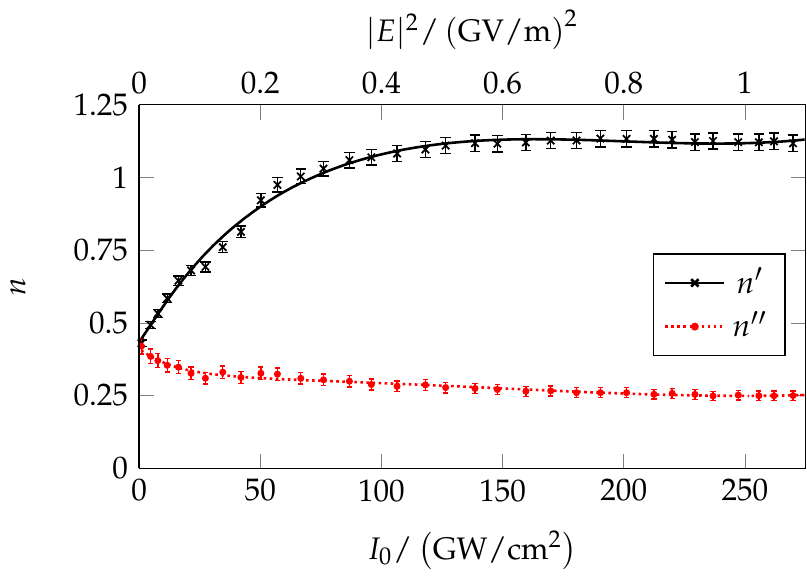}
\caption{Despite the saturation behavior at high intensities, we can
correctly fit both the real and imaginary parts of the index refraction
of ITO for intensities up to 275~GW/cm$^{2}$ with the addition of
appropriate $\chi_{\textrm{}}^{(5)}$ and $\chi^{(7)}$ terms.}

\centering{}\label{Fig: simultaneousfit}
\end{figure}

%We extract $\epsilon$ from Eq.~(\ref{eq:PNL power series}):
%\begin{align}
%\epsilon & =1+\sum_{j\,\textrm{odd}}^{\infty}c_{j}\chi^{(j)}|E|^{j-1},\label{eq:higher order epsilon}
%\end{align}
We extract $\epsilon$ from Eq.~(\ref{eq:PNL power series}) and fit to the real and imaginary parts of the intensity-dependent
refractive index of ITO for intensities up 275~GW/cm$^{2}$.
The resulting curve correctly describes both $n'$ and $n''$ at all intensities (Fig.~\ref{Fig: simultaneousfit}). The extracted
$\chi^{(j)}$ values are listed in Table~\ref{TAB:extracted values}.
The real part of $n_2$ %the nonlinear index 
calculated using $\chi^{(3)}$
extracted in this process is $\textrm{Re\,}(n_{2})=\nobreak0.016\textrm{\,cm}^{2}/\textrm{GW}$,
in agreement with the previous measurement. 

We use Pearson's statistical chi-squared value to determine the most appropriate fit for this data set~\cite{Young2014}.
Fits with fewer nonlinear orders than $\chi^{(7)}$ 
yield a significantly larger statistical error (\emph{i.e.,} larger chi-squared). Including orders beyond $\chi^{(7)}$
only improves the statistical error marginally, indicative
of overfitting. We thus attribute the nonlinear refraction of ITO
to $\chi^{(3)}$, $\chi^{(5)}$ and $\chi^{(7)}$ nonlinearities at the investigated intensities.

\begin{figure}[htb]
\centering{}
\includegraphics[width=0.9\columnwidth]{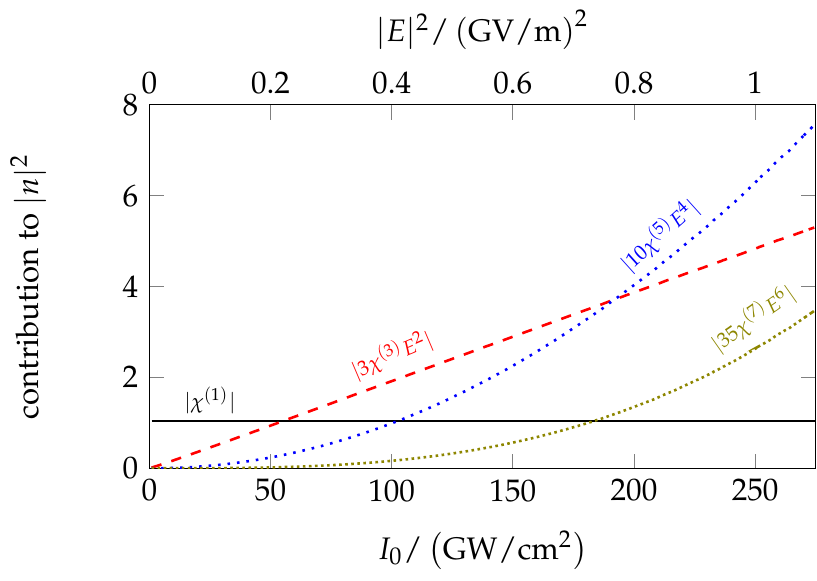}
\caption{The absolute contribution of the various orders of the nonlinear susceptibility
to the refractive index of ITO at the wavelength where $\textrm{Re}(\epsilon^{(1)})=0$.
These contributions are estimated using the values for $\chi^{(j)}$
in Table~\ref{TAB:extracted values}.} %At the highest intensities, the linear contribution to the index of refraction is dwarfed by all the nonlinear contributions up to $\chi^{(7)}$.}
\label{Fig: contribution}
\end{figure}

To examine the contributions of different orders to the refractive
index, we plot them as functions of intensity in Fig.~\ref{Fig: contribution}.
At the highest input intensities,
the linear refractive index makes only the fourth largest contribution
to the total refractive index, providing further evidence that nonlinear
optical effects cannot be treated solely as perturbations to linear
optics. In fact, for an accessible range of operating intensities,
nonlinear effects dominate the optical response of this material.

At the maximum utilized pump intensity $I_{\textrm{max}}\nobreak=\nobreak275\textrm{\,GW/cm\ensuremath{^{2}}}$,
the $\chi^{(5)}$ term makes
the largest contribution to the total refractive index; this term
contributes more than the $\chi^{(3)}$ term, which is typically considered
to be the dominant mechanism for $n_{2}$ and $\Delta n$. Additionally,
the $\chi^{(7)}$ term is also significant, accounting for 20\% of
the total susceptibility. %In Table~\ref{TAB:extracted values}, we list the magnitude of their contributions to the refractive index at the maximum intensity.

Next, we use these extracted values to directly address whether the
large nonlinearities that have been observed in ENZ materials violate
traditional formulations of nonlinear optics that are based on the
power series expansion of the nonlinear polarization.

The convergence of the power series in Eq.~(\ref{eq:PNL power series}) can be
determined using the ratio test:
%\begin{equation}
%\lim_{j\rightarrow\infty}\left|\frac{a_{j+1}}{a_{j}}\right|<1.\label{eq:converging criterion generic}
%\end{equation}
%Applying this criterion to the series yields
\begin{equation}
\lim_{j\rightarrow\infty}\left|\frac{c_{j+2}\chi^{(j+2)}|E|{}^{j+1}}{c_{j}\chi^{(j)}|E|{}^{j-1}}\right|<1.\label{eq:converging criterion}
\end{equation}
When this inequality is satisfied, the series is said to be converging. 
Thus, we see that $\Delta n/n$ and $n_{2}$ are not the relevant
quantities for a discussion on convergence of the nonlinear polarization,
even though their magnitude is critical to the convergence of Eq.~(\ref{eq:n_power_series}).
Instead, the various nonlinear orders of the susceptibility $\chi^{(j)}$
determine its convergence. We note from Fig.~\ref{Fig: contribution} 
that the first few terms in the series violate the inequality, since
at the maximum intensity investigated the fifth order contribution
to refraction ($7.6\pm0.3$) is larger than the third order contribution ($5.30\pm0.09$),
which in turn is larger than the linear contribution ($1.043\pm0.004$). However,
the corresponding ratio between the $\chi^{(7)}$ and $\chi^{(5)}$
terms obeys the criterion in Eq.~(\ref{eq:converging criterion}). 

\begin{table}[htb]
\centering{}%
\begin{tabular}{|c|c|c|c|} \hline 
$j$ & $\textrm{Re}\,\chi^{(j)}/(10^{-9}\textrm{m}/\textrm{V})^{j-1}$ & $\textrm{Im\,}\chi^{(j)}/(10^{-9}\textrm{m}/\textrm{V})^{j-1}$ \tabularnewline \hline % & $|c_{j}\chi^{(j)}E_{\textrm{max}}^{j-1}|$\tabularnewline \hline 
$1$ & $-0.980\pm0.008$ & $0.36\pm0.01$ \tabularnewline \hline % & $1.043\pm0.004$\tabularnewline \hline 
3 & $1.60\pm0.03$ & $0.50\pm0.05$ \tabularnewline \hline % & $5.30\pm0.09$\tabularnewline \hline 
5 & $-0.63\pm0.02$ & $-0.25\pm0.04$ \tabularnewline \hline % & $7.6\pm0.3$\tabularnewline \hline 
7 & $(7.7\pm0.3)\times10^{-2}$ & $(3.5\pm0.8)\times10^{-2}$ \tabularnewline \hline % & $3.5\pm0.2$\tabularnewline \hline 
\end{tabular}\caption{Values extracted from the fit to Eq.~(\ref{eq:PNL power series}) %(\ref{eq:higher order epsilon})
with a third, fifth and seventh-order nonlinearity.} %The last column lists the varying contributions to refraction at $I_{\textrm{max}}=275\,\textrm{GW}/\textrm{cm}^{2}$ ($|E_{\textrm{max}}|=1.0\times10^{9}\textrm{\,V}/\textrm{m}$), the highest intensity investigated.}
\label{TAB:extracted values}
\end{table}

Though we have no access to the coefficients in the limit of $j\rightarrow\infty$,
we remark that they must be negligible at the investigated intensities
since we can accurately fit to the refraction without them. For example,
$\chi^{(9)}$ was not found to be statistically different from zero;
therefore, its contribution to the refractive index is insignificant
even at the maximum intensity investigated, and in particular must
be smaller than the seventh-order term, obeying the convergence criterion.
Thus, we conclude that the large nonlinear index of refraction that
is observed in ENZ materials is nonetheless consistent with a power
series description of the nonlinear polarization.

%\section{Discussion and conclusion}
The above treatment and discussion prominently demonstrate that there
is indeed a need to reinterpret established quantities
related to the optical response %(such as $n_2$)
in materials with small indices of
refraction. We conclude that, in this unique scenario, it is no longer
appropriate to use the approximation of the intensity-dependent
index of refraction that only depends linearly on the intensity, even
when only accounting for $\chi^{(3)}$ nonlinearities. Instead, we
have introduced a more general equation with a square root dependence.
The linear slope with which the community is familiar is merely a special case of Eq.~({\ref{eq:intensity-dependent index withchi}}) for when the linear index is large. Because it is based on so few assumptions, our method
will continue to work in cases that are not explicitly considered
in this letter, such as if $|n_{0}|\gg0$ or $|\Delta n/n|\sim1$.
The generalized equation developed here has the benefit of preserving	
the standard historical definition of $n_{2}$ as a function of $\chi^{(3)}$
(Eq.~(\ref{eq:n2_definition})) as well as the physical definition
of $n_{2}$ as the \emph{initial} slope for the refractive index with
respect to the applied optical intensity, \emph{i.e.,}
%\begin{equation}
$n_{2}\equiv\lim_{I\rightarrow0}\partial n/\partial I$.
%\end{equation}
However, since the definition of $n_{2}$ is problematic in the context
of low-index materials, $\chi^{(3)}$ or $\Delta n$ should be used
to characterize materials, instead.

We have demonstrated how to extend our generalized equation to incorporate
higher-order nonlinear terms and absorption. Besides the assumption
that the nonlinear susceptibility can be expanded in a power series,
this treatment tracks the measured refraction for intensities up 275~GW/cm$^{2}$ without the need for
a detailed microscopic model or empirical saturation equations~\cite{Smektala2000,Chen2006%,Fontana2016\tabularnewline
}. 
Though our treatment cannot make predictions for even higher intensities, it enables quantitative statements regarding the convergence of the material polarization.
It may also be used to systematically
estimate the magnitude of higher-order contributions. Incidentally,
we have shown that the nonlinear properties of ITO are even more striking
than previously realized. At the highest probed intensities, the index
of refraction is dominated by a fifth-order nonlinearity whose contribution
grows roughly with $I^{2}$. We have also detected significant contributions
to refraction caused by seventh-order nonlinearities. The nonlinear
contributions from $\chi^{(3)}$, $\chi^{(5)}$ and $\chi^{(7)}$
terms each exceed the linear refraction term, making ENZ materials
\textemdash{} to the best of our knowledge \textemdash{} the first
solid-state platform to possess this property. 

Finally, we have quantitatively shown that there is no evidence that
the power series expansion for the nonlinear polarization in ENZ materials
diverges at the wavelength where the linear permittivity vanishes.
However, the dominant higher-order nonlinear contributions that have
been observed reveal that ENZ materials operate in a regime where
nonlinear optical effects can no longer be treated as a perturbation
to linear optics.

% Bibliography

%\bibliography{library}

% Full bibliography added automatically for Optics Letters submissions
% Note that this extra page will not count against page length

\clearpage
%\bibliographyfullrefs{library}
 
%Manual citation list

\end{document}